\documentclass[10pt,aps,pre,showpacs,twocolumn,superscriptaddress,longbibliography,amsmath,amssymb]{revtex4-2}
\usepackage[pdftex]{graphicx}
\usepackage{color}
\usepackage[pdftex,bookmarks,colorlinks,breaklinks]{hyperref} 
\hypersetup{linkcolor=red,citecolor=blue,filecolor=dullmagenta,urlcolor=blue} 
\hypersetup{pdftoolbar=true,pdfpagemode=UseNone,pdfstartview=FitH, colorlinks=true,extension=}
\newcommand{\vol}{\text{Vol}(\Omega)}

\renewcommand{\Re}{\mathrm{Re}}
\renewcommand{\Im}{\mathrm{Im}}

\makeatletter
\providecommand{\hateq}{\mathrel{\mathpalette\my@hat@eq\relax}}
\newcommand{\my@hat@eq}[2]{%
	\begingroup
	\sbox\z@{$\m@th#1=$}%
	\ooalign{%
		\hidewidth\raisebox{-0.3\ht\z@}{$\m@th#1\widehat{}$}\hidewidth\cr
		\box\z@\cr
	}%
	\endgroup
}
\graphicspath{{./figs/},{./}}

\newcommand{\erm}{\textrm{e}}
\begin{document}
\title{Non-Weyl Behavior Induced by Superradiance: A Microwave Graph Study}

\author{Junjie Lu}
\affiliation{Universit\'{e} C\^{o}te d'Azur, CNRS, Institut de Physique de Nice (INPHYNI), 06108 Nice, France, EU}
\author{Tobias Hofmann}
\affiliation{Fachbereich Physik der Philipps-Universit\"{a}t Marburg, D-35032 Marburg, Germany, EU}
\author{Hans-J\"{u}rgen St\"{o}ckmann}
\affiliation{Fachbereich Physik der Philipps-Universit\"{a}t Marburg, D-35032 Marburg, Germany, EU}
\author{Ulrich Kuhl}
\email{ulrich.kuhl@univ-cotedazur.fr}
\affiliation{Universit\'{e} C\^{o}te d'Azur, CNRS, Institut de Physique de Nice (INPHYNI), 06108 Nice, France, EU}
\affiliation{Fachbereich Physik der Philipps-Universit\"{a}t Marburg, D-35032 Marburg, Germany, EU}

\date{\today}

\begin{abstract}
    We study experimentally the manifestation of non-Weyl graph behavior in open systems using microwave networks.
    For this a coupling variation to the network is necessary, which was out of reach till now.
    The coupling to the environment is changed by indirectly varying the boundary condition at the coupling vertex from Dirichlet to Neumann using a dangling bond with variable length attached the coupling vertex.
    A transformation of equal length spectra to equal reflection phase spectra of the dangling bond allows to create spectra with different fixed coupling strength.
    This allows to follow the resonances in the complex plane as a function of the coupling.
	While going from closed (Dirichlet) to fully open (Neumann) graph we see resonances escaping via a superradiant transition leading to non-Weyl behavior if the coupling to the outside is balanced.
	The open tetrahedral graph displays a rich parametric dynamic of the resonances in the complex plane presenting loops, regions of connected resonances and resonances approaching infinite imaginary parts.
\end{abstract}

\maketitle

\begin{figure}
	\includegraphics[width=\columnwidth]{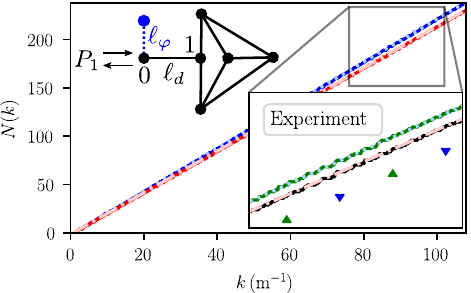}
	\caption{\label{fig:Nk}
		The counting functions $N(k)$ for the tetrahedral graph with additional dangling bond $\ell_d$ (see the black sketch) is presented, where the boundary condition at vertex 0 is Neumann (perfect coupling: $\varphi=0$) in blue or Dirichlet (no coupling $\varphi=\pi$) in red.
		The filled circles in the sketch indicate vertices with Neumann boundary.
		Additionally, the theoretical $N(k)$ from Eq.~(\ref{eq:Weyl}) is shown, where $L\!=\!L_{tot}=7.001$\,m (non-balanced, dotted line) and $L=L_{red}\!=\!L_{tot}-\ell_d\!=\!6.757$\,m (balanced, dashed-dotted line) and in both cases $c=-2.5$.
		The experimental data in the range $k\in [76,95]$\,m$^{-1}$ is presented in the zoom, where all resonances could be extracted.
		The green step function corresponds to $\varphi=1.34$ and the black to $\varphi=0$.
		The triangles indicate the position of $k$-values related to the ``resonances with infinite width'' leading to the difference in slopes in this range.
		The two blue down and two green up triangles mark the wave numbers at $(n+1/4) k_d$ and $(n+3/4) k_d$ (with $k_d=\pi/\ell_d$), respectively, indicating the difference of two resonances inside the presented range between the two graphs (for more details see text).
	}
\end{figure}

\emph{Introduction --}
One of the famous equations in wave systems is Weyl's law, which have been first used to explain black body radiation\cite{wey12a,wey12b,wey12c}.
It counts the number of eigenvalues smaller than the energy $E=k^2$ in quantum systems, or in general wave systems smaller than a wave number $k$, in average.
In a $d$-dimensional domain $\Omega$ it is given by:
\begin{equation}\label{eq:Weyl}
    \begin{matrix}
        N(k) &=& \frac{\omega_d \vol }{(2\pi)^d}k^d + O(k^{d-1})\\
        &\overset{\raisebox{0.3ex}{d=1}}{\longrightarrow}&\frac{L}{\pi} k + c + O(k^{-1})\,,
    \end{matrix}
\end{equation}
where volume $\vol$ is the volume and $\omega_d$ is the volume of the $d$-dimensional unit sphere.
The corrections have been subject intensive research\cite{ivr16,odz19}, as well as the extension of the Weyl law to vectorial waves\cite{bal71b}.
The last part is for the 1D case of a graph with total bond length $L$, which will be the system of interest in this paper (see sketch in Fig.~\ref{fig:Nk}).
The constant $c=O(1)$ in the large $k$ limit depends on the number of the vertices and the boundary condition (BC) at the vertices\cite{gnu06b}.

New features come into play if the system is gradually opened.
Now one has to count resonances as well as eigenvalues (which stem from bound states in the continuum (BICs)\cite{hsu16,law21,wan23}) and the definition of the volume becomes questionable as the system is coupled to an environment.
For sufficiently weak coupling, Weyl's law is still working well as long as the resonances acquire only a small negative imaginary part $\Im(k)$.
Once the coupling to the outside gets strong effects like the fractal Weyl law for chaotic systems can be found, where $N(k)\propto k^d$, and $d$ is no longer the spatial dimension but is related the dimension of the classical repeller of the system\cite{zwo89b,non05,kop10}.
For experimental realizations see Refs.~\cite{lu03,pot12}.

Here we shall concentrate on graphs or networks (see sketch in Fig.~\ref{fig:Nk}), which are conceptionally simple systems but can show statistical properties described by random matrix theory, if the graph is sufficiently complex\cite{gnu04b,plu13a,plu14}.
A graph consists of vertices connected by bonds.
The wave function within each bond is obtained from a wave equation, in case of quantum graphs the Schrödinger equation, in electromagnetic networks the Helmholtz equation\cite{hul04}.
At the vertices the wave has to obey BCs corresponding to generalized Kirchhoff laws.
Graphs have been subject to intensive research in mathematics\cite{ber03} and physics\cite{kot97a,kot01,gnu04b}.
An overview can be found in \cite{ber13,ber17}.
Graphs can be implemented as microwave networks, where they have been used to study experimentally the three different Gaussian ensembles the orthogonal\cite{hul04}, the unitary\cite{law17,reh18} and the symplectic one\cite{reh16,reh18,lu20}, even though, due to the small number of bonds and vertices typically used, there are deviations from Gaussian behavior\cite{hof21}.
The high flexibility of graphs allowed to perform detailed experimental studies on properties of the closed graph, e.g., isospectrality\cite{hul12,law14a}, as well as in open graphs on scattering matrices or Wigner reaction matrix elements\cite{law10}, absorption\cite{hul05a,all14a}, coherent perfect absorbers\cite{che20} or Wigner time delays\cite{che21a,che21b}.

One of the shortcomings of previous graph experiments is the lack of adjustable couplings to the environment.
It is exactly this lack which we overcome in this paper and use it to determine the characteristic of non-Weyl behavior by following the resonances in the complex plan while varying the coupling.

Opening the graph on vertices can produce non-Weyl behavior in the sense that not the full length determines the slope of the Weyl law but a reduced length $L'<L$\cite{dav10,dav11,exn11,lip15,lip16,exn20}.
This can happen if vertices are \emph{balanced}, i.e., the number of semi-infinite leads connected to the vertex that are coupling to the environment are the same as the number of internal bonds, and the standard Kirchhoff conditions are met at this vertex.
Figure~\ref{fig:Nk} presents numerical simulations for a balanced (Neumann BC, blue step function) and a non-balanced (Dirichlet BC, red step function) graph (see sketch in black).
The simulations have been done using the bond matrix approach \cite{kot03}.
In the presented case of a vertex with valency of two the non-Weyl behavior is seen, but for other valencies as, e.g., presented in Ref.~\cite{law19b}, the situation is more complicated.

In the present case the total size $L=L_{tot}$ has to be taken into account in Weyl's law (see dotted line) if scattering is taking place at vertex 0, but if the vertex is balanced there is no scattering at vertex 0, thus reducing the length entering the Weyl formula by $\ell_d$ ($L=L_{tot}-\ell_d$, see dashed-dotted line).
The graph presented in this paper consists of a dangling bond with length $\ell_d=0.2435$\,m connecting vertex 0 to vertex 1 and a tetrahedron attached at vertex 1 to this dangling bond with lengths $\ell_{1,\dots,7}\,\mathrm{m}^{-1}$=0.949, 0.374, 1.75, 1.59, 0.868, 0.786, and 0.438.
In case of the experiment the lengths correspond to electrical lengths.
$P_1$ represent a semi-infinite lead corresponding in the experiment to port 1 of a vector network analyzer (Agilent 8720ES), measuring the reflection amplitude $S_{11}$.
Each vertex (black circle) is a T-junction obeying Kirchhoff's law, i.e., Neumann BC.
Fitting the shown $\Re(k)$ range to the Weyl law we obtained $L=7.001 \pm 0.002$\,m and $L'=6.758 \pm 0.006$\,m in pefect agreement with the experimental $L=7.0009$\,m and $L'=6.757$\,m, respectively.
Note, that, since the reference point of the scattering is still at vertex 0, the length $\ell_d$ still appears in the reflection amplitude an additional acquired global phase $\Phi_d=2 k \ell_d$.

Non-Weyl behavior has been observed experimentally by {\L}awniczak et~al.~\cite{law19b} for a vertex with two semi-infinite leads and two internal bonds.
In a small wavenumber region the authors noticed a reduction of the number of resonances by two or three, in good agreement to the theoretical expectations.
In the experiment the loss of resonances\cite{law17} due to negligible coupling to the openings (or no coupling as in case of BICs) is always a problem.
In particular, there is never a guarantee that all resonances are found.
This problem is overcome in the present paper by following the resonances parametrically in the complex plane towards the limit of ``infinite width'' thus monitoring the superradiant transition to non-Weyl behavior.

The variation of the coupling is realized by changing the BC at vertex 0.

A variation of the coupling parameter allows to study resonance trapping\cite{mag99,per98a,per00,stoe02c} and superradiant or doorway states\cite{aue11,guh09,per99b,wei21} as well as non-Weyl graphs\cite{dav10,dav11,exn11,lip16} analytically, numerically as well as experimentally.

\emph{Experimental setup --}
It is not trivial to implement experimentally a variable BC or variable coupling constant over a large wave number range.
This problem is solved by means of a T-junction, where one port is connected to the analyzer, the second port to a dangling bond with length $\ell_d$ and the third one to a bond with length $\ell_\varphi$ open terminated (corresponding to Neumann BC) (see blue part of the sketch in Fig.~\ref{fig:Nk}).
The bond length $\ell_\varphi$ is varied using a phase shifter (ATM, P1507), and reflection amplitudes $S_{11}(k,\ell_\varphi)$ have been measured for each length.

\begin{figure}
	\centering
	\includegraphics[width=.99\linewidth]{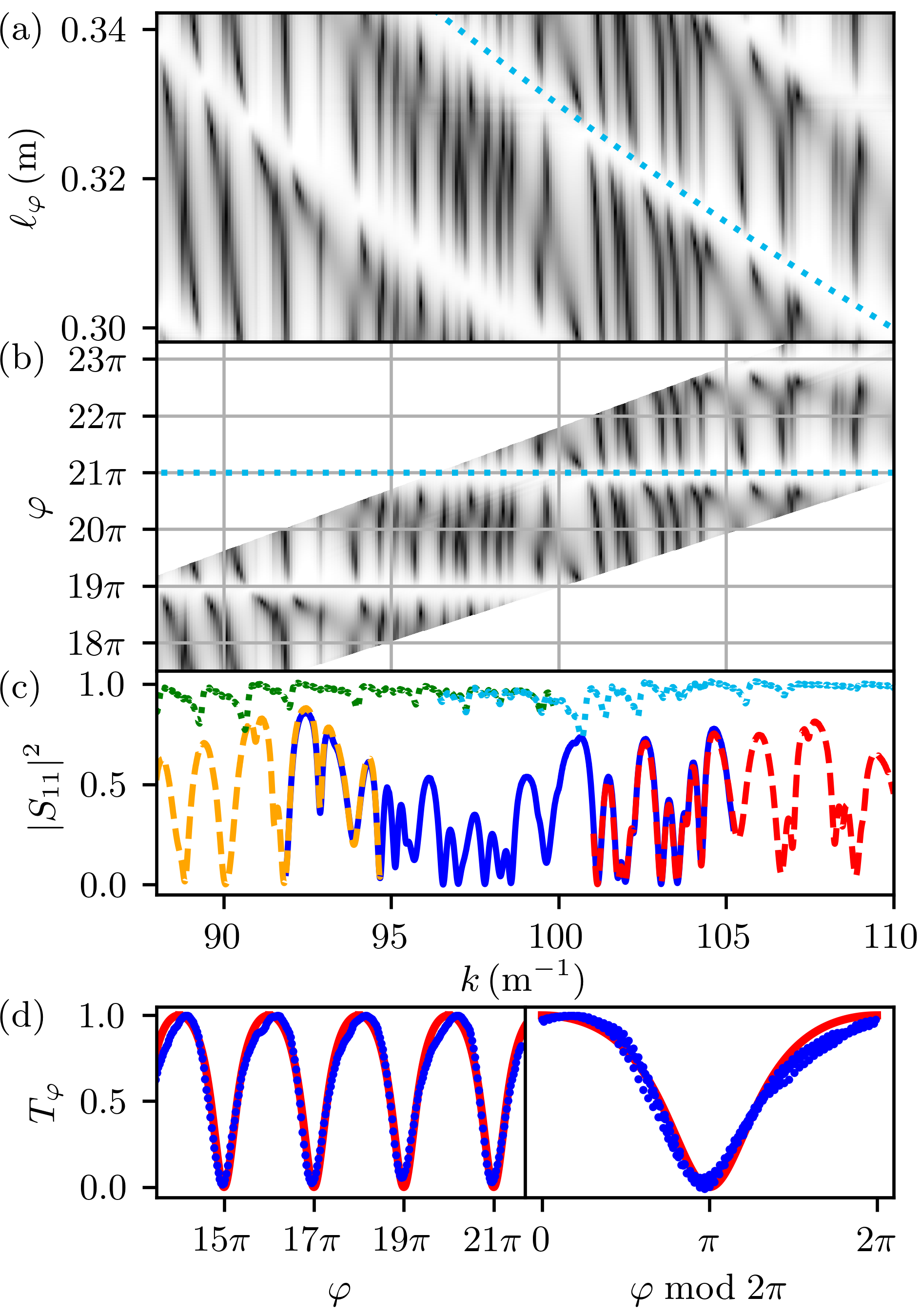}
	\caption{\label{fig:transform_l2phi}
		(a) Reflection $\left| S_{11}\right| ^2$ in dependence of wave number for constant $\ell_\varphi$ in a gray scale, where black corresponds to 0 and white to 1 in a linear scale.
		The measurements for different $\ell_\varphi$ are stacked onto each other.
		For the cyan dotted line the phase $\varphi=2\ell_\varphi k$ corresponds to $21\pi$.
		(b) The same data but now rearranged to constant $\varphi$ using Eq.~(\ref{eq:varphi}).
		(c) Superposition of the reflection spectra shown in (b) along the lines $\varphi=n\pi$ for $n=18$ (dashed orange), $19\pi$ (dotted green), $20\pi$ (solid blue), $21\pi$ (dotted cyan), and $22\pi$ (dashed red).
		Odd $n$ values correspond to Neumann, even $n$ values to Dirichlet boundary conditions at the coupling T-junction.
		(d) Transmission intensity through the coupling T-junction.
		The blue dots have been obtained from $1-|\langle S_{11}(k,\varphi)\rangle_k|^2$ and the red line corresponds to Eq.~(\ref{eq:T_varphi}).
	}
\end{figure}

Figure~\ref{fig:transform_l2phi}(a) shows the reflection intensities $|S_{11}|^2$ in a gray scale plot.
$|S_{11}|^2$ is maximal along lines given by $2 k \ell_\varphi= (2n+1)\pi$ ($n$ integer), one of them is indicated as cyan dotted line ($2n+1=21$).
This can be understood from the properties of the coupling T-junction:
The bond $\ell_\varphi$ will reflect the wave with amplitude $r_\varphi=\exp(i\varphi)$ back to the vertex 0, where 
\begin{equation}\label{eq:varphi}
	\varphi=2 k \ell_\varphi\,.
\end{equation} 
For $\varphi=(2n+1)\pi$ the amplitude is inverted ($r_\varphi=-1$) and the wave at vertex 0 has to be zero imposing Dirichlet BC, thus the incoming wave is totally reflected at vertex 0.
For $\varphi=2n\pi$ we obtain Neumann BC at vertex 0 and the wave is propagating through the vertex without any scattering, i.e., perfect matching.
Varying $\varphi$ from 0, via $\pi$ to $2\pi$ thus means a variation of the BC from Neumann to Dirichlet back to Neumann with a corresponding variation of the coupling to the system.

Transforming the spectra taken for constant $\ell_\varphi$, shown in Fig.~\ref{fig:transform_l2phi}(a) to spectra for constant $\varphi$ using Eq.~(\ref{eq:varphi}) we obtain the spectra shown in Fig.~\ref{fig:transform_l2phi}(b).
The empty triangular regions in the upper left and lower right present the fact that due to the limitations of the phase shifter the length variation of $\ell_\varphi$ is limited to a maximum of 4\,cm.
The same technique has been used to realize graphs with GSE symmetry\cite{reh16,reh18,lu20,law24}.
In Fig.~\ref{fig:transform_l2phi}(c) the reflection spectra for the integer $\pi$ values for constant $\varphi=n\pi$ are combined.
For even $n$ the resonances have large imaginary parts and for odd $n$ they are small.
For odd $n$ there should be a perfect reflection of 1, which is not the case as we neglected effects of absorption in the bond $\ell_\varphi$, leading to a reduction of the reflection to $|r_\varphi|^2<1$.
Note that in the regimes where contributions from different $n$ overlap the spectra are nicely matching illustrating the precision of the experimental setup.

Starting from the scattering matrix of a T-junction and imposing the condition that the back scattering amplitude from the bond $\ell_\varphi$ is given by $\exp(i\varphi)$ the transmission amplitude $t_\varphi$ through vertex 0 is given by 
\begin{equation} \label{eq:T_varphi}
t_\varphi = \frac{2 (1+\erm^{i \varphi})}{3+\erm^{i \varphi}}\,.
\end{equation} 
In Fig.~\ref{fig:transform_l2phi}(d) the transmission intensity $T_\varphi=|t_\varphi|^2$ as calculated from Eq.~(\ref{eq:T_varphi}) is shown by red solid lines.
The blue dots correspond to $1-|\langle S_{11}(k,\varphi)\rangle_k|^2=T_\varphi$, where $\langle . \rangle_k$ denotes an average over the measured wave number for constant $\varphi$.
A good agreement is found.

\emph{Experimental results --}
From the measured reflection spectra $S_{11}(k,\varphi)$ the resonances are extracted applying the harmonic inversion technique\cite{kuh08b}.
In the zoom of Fig.~\ref{fig:Nk} the step function obtained for $\varphi=1.34$ (green) and $\varphi=0$ (black) in the wave number range of $k\in [76,95]$\,m$^{-1}$ are presented.
A good agreement with the theoretical predictions (dotted and dashed-dotted line, Eq.~(\ref{eq:Weyl}) with $L=7.001$\,m and $L=6.757$\,m, respectively) is found.
As we did not have access to all resonances below $k=k_{min}=75$\,m$^{-1}$, $N(k_{min})$ was adjusted.
Fitting the experimental step function to the Weyl law gives $L=7.01\pm0.02$\,m and $L'= 6.76\pm0.02$\,m in good agreement with the prediction.
The two $N(k)$ differ in this range by 2 as expected from theory.
Experimentally it is extremely difficult to find all resonances.
Just below the shown range at around $k$\,=\,94.5\,m$^{-1}$, e.\,g., we were not able to detect one resonance due to small coupling.
Even in the numerical simulation this resonance was hard to detect from the reflection amplitude.
This situation can be improved by applying a parametric variation.
The most natural parameter in this context is $\varphi$.
The parametric dynamics will allow us in particular to identify the resonances escaping to $-i\infty$ thus leading to a reduction of the prefactor in the Weyl law for the balanced case.

\begin{figure*}
	\includegraphics[width=\linewidth]{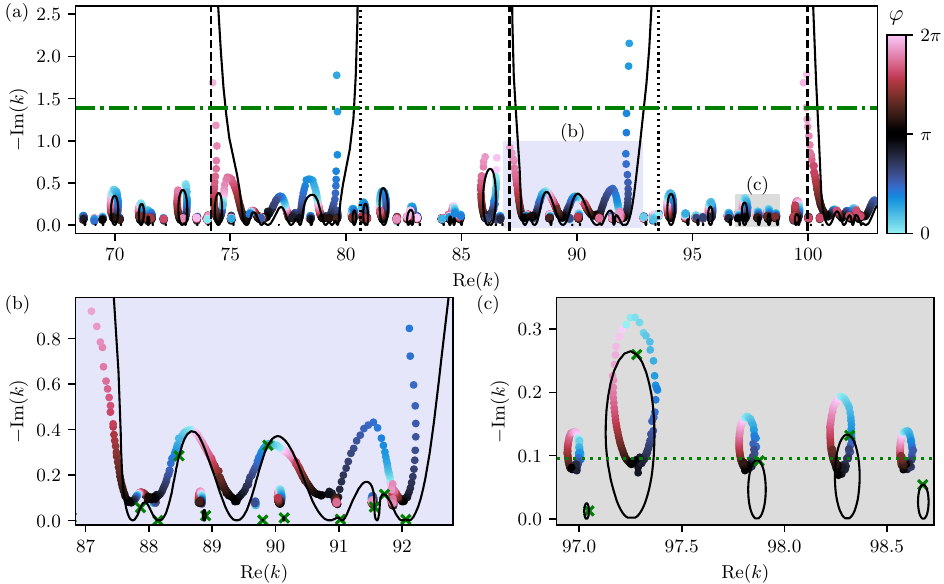}
	\caption{\label{fig:reso_tetrahedron}
		(a)~The colored circle display the poles extracted from the experimental data by harmonic inversion.
		The color of the circle indicates the $\varphi$ value of the measured spectra (value see colorbar).
		The solid black lines correspond to poles calculated via numeric simulation not including absorption.
		The dotted and dashed vertical lines correspond to $(n+1/4) k_d$ and $(n+3/4) k_d$ where $k_d=\pi/\ell_d \approx 12.90\,\textrm{m}^{-1}$, respectively.
		The dashed dotted green line is indicating $\log(3/\ell_{min})$.
		(b) and (c) are enlargements of the correspondingly marked regions in (a).
		The green crosses correspond to the resonances of the open tetrahedral graph without dangling bond $\ell_d$.
		The dotted horizontal line shows the estimated global absorption $\gamma=0.0952$\,m$^{-1}$.
	}
\end{figure*}

Figure~\ref{fig:reso_tetrahedron} shows the resonances extracted from the experimentally measured reflection $S_{11}(k,\varphi)$ by means of the harmonic inversion (colored circles) for the tetrahedral graph (see inset of Fig.~\ref{fig:Nk}) in the complex $k$-plane and the parametric dependence on $\varphi$ (color scale on the right).
The resonances separate into two different kind of wave number ranges:
(i) from $(n+1/4) k_d$ (dotted lines) to $(n+3/4) k_d$ (dashed lines), $k_d=\pi/\ell_d \approx 12.90\,\textrm{m}^{-1}$, only loops are observed, and 
(ii) from $(n+3/4) k_d$ (dashed lines) to $(n+5/4) k_d$ (dotted lines)
the resonances show a continuous behavior with some additional small loop structures.
When approaching $\varphi \to 0^+$ (blue dots) from above a resonance becomes very broad close to $(n+1/4) k_d$, whereas for $\varphi \to 2\pi^- \hateq 0^-$ (red dots) this happens at $(n+3/4) k_d$.
These resonances correspond to superradiant states disappearing in the continuum whenever the perfect matching condition holds, i.e., whenever the vertex is balanced.
In the zoom in Fig.~\ref{fig:Nk} these states are indicated by blue down and green up triangles, respectively.
Thus, it is this disappearance of the superradiant states being responsible for the reduction of the length $L$ in Weyl's law.
Experimentally there is a limit to resolve the resonances using the harmonic inversion which in our case is at about $-\Im(k)=1.5$\,m$^{-1}$.
This is in the spirit of fractal Weyl law where the counting of resonances in open systems is restricted to a finite strip below the real axis\cite{non05,pot12}.
The horizontal dashed dotted line in Fig.~\ref{fig:reso_tetrahedron}(a) corresponds to $-\Im(k)=\log(3/\ell_{min})$, where $\ell_{min}$ is the minimal length in the tetrahedron and 3 is the maximal number of bonds connected to a vertex, a prediction obtained in \cite{ing22} for non-balanced graphs with Neumann BCs only.
All resonances apart from the superradiant ones are below this limit.

In addition, we calculated the poles from the bond scattering matrix of the graph\cite{kot03}, where the bond $\ell_\varphi$ was taken into account assuming a zero length and a fixed phase $\varphi$.
To get a good agreement all lengths $\ell_i$ and the global absorption $\gamma$ had been matched by an optimization procedure reducing the difference of the measured complex reflection starting from the known electrical length of the cables.
The variation of the $\ell_i$ was of the order of a few mm, within the experimental uncertainty of the cable lengths.
A complex wave number $k=k_r+i\gamma$ was used in the numerical calculations to take care of absorption, resulting in $\gamma=0.0952$\,m$^{-1}$.
The solid lines correspond to this simulation for $\gamma=0$, i.e., without including absorption.
The shift in the imaginary part of the loops observed in Fig.~\ref{fig:reso_tetrahedron}(c) corresponds to the global absorption obtained (horizontal dotted line).
We verified that the poles calculated from the scattering matrix touch the real axis for $\varphi=\pi$.
In this case the poles correspond to the eigenvalues of the closed graph with a Dirichlet condition at vertex 0.
For $\varphi=0$ the resonances correspond to the spectra of the tetrahedron without the dangling bond [crosses in Fig.~\ref{fig:reso_tetrahedron}(b) and (c)], i.e., Neumann condition with perfect matching at vertex 0.
Note that the maximal $-\Im(k)$ for the individual resonance is not necessarily reached for $\varphi=0$ or $2\pi$, but can be higher at other $\varphi$ values.
Looking into the loop structure we observe that while increasing from $\varphi=\pi$ to $2\pi$ first the resonance width increases, then attains the maximum and decreasing while approaching $\pi$ from 0.
Similar features are found for resonance trapping but there the resonances do not return to its original real parts $\Re(k)$\cite{mag99,per98a,per00,rot01,kot04,stoe02c}.
Furthermore we observe that in the range from $(n+3/4) k_d$ (dashed vertical lines) to $(n+1+1/4) k_d$ (dotted vertical lines) the two different kind of superradiant states occurring slightly below $2\pi$ and above 0 are connected by a line in dependence of the parameter.
On the other hand in the range from the resonances $(n+1/4) k_d$ to $(n+3/4) k_d$ form loops only.

\emph{Conclusion --}
We implemented experimentally a variation of the BC at the coupling vertex in a graph using a T-junction coupling, where at one port of the T-junction a bond with an reflecting end is attached.
This parametric variation allows to study superradiant states and the emergence of non-Weyl.
The variation of the BC can be performed also on vertices with higher valencies as well as on vertices not coupled to the outside.
According to the interlacing theorem\cite{ber13,ber17} the spectra of closed graphs differing in the BC in only one vertex are strictly alternating.
The technique developed in this work allows to study this behavior experimentally.
Also it provides an additional parameter for studies of coherent perfect absorption and exceptional points.

\begin{acknowledgments}
J.L.~acknowledges financial support from the China Scholarship Council via Grant No.~202006180008 and from the university of Marburg for a three month visit at the quantum chaos group in Marburg.
\end{acknowledgments}

\end{document}